\documentclass[fleqn,usenatbib]{mnras}

\usepackage[T1]{fontenc}
\usepackage{ae,aecompl}

\usepackage{graphicx}	
\usepackage{amsmath}	
\usepackage{amssymb}	
\usepackage{tabularx,booktabs,tablefootnote}
\usepackage{multirow}
\usepackage{hyperref}
\usepackage{newtxtext,newtxmath}

\newcommand{\eu}[5]{\mbox{$#1\,^#2{\rm #3}^{#4}_{\rm #5}$}}
\newcommand{\vmicro}{$\xi_{\rm t}$}
\newcommand{\kms}{km\,s$^{-1}$}
\newcommand{\te}{$T_{\rm eff}$}
\newcommand{\logg}{$\log{g}$}
\newcommand{\vsini}{$v\sin{i}$}
\newcommand{\vmacro}{$\zeta_{\rm RT}$}

\newcommand{\vald}{\textsc{VALD3}}

\newcommand{\SED}{\textsc{SED}}

\title[Ap star HD~152564]{Doppler imaging of a southern ApSi star HD~152564}

\author[I. Potravnov et al.]{
I. Potravnov$^{1}$\thanks{E-mails: ilya.astro@gmail.com (IP), ryabchik@inasan.ru (TR)},
T. Ryabchikova$^{1}$\footnotemark[1],
N. Piskunov$^{2}$,
Y. Pakhomov$^{1}$,
A. Kniazev$^{3,4,5,6}$
\\
$^{1}$Institute of Astronomy of RAS, Pyatnitskaya str., 48, 119017, Moscow, Russia\\
$^{2}$Department of Physics and Astronomy, Division of Astronomy and Space Physics, Uppsala University, P.O. Box 516, 751 20, Uppsala, Sweden\\
$^{3}$South African Astronomical Observatory, PO Box 9, 7935 Observatory, Cape Town, South Africa\\
$^{4}$Southern African Large Telescope, Cape Town, 7935, South Africa\\
$^{5}$Sternberg Astronomical Institute, Lomonosov Moscow State University, Moscow, Russia\\
$^{6}$Special Astrophysical Observatory of RAS, Nizhnij Arkhyz, Karachai-Circassia 369167, Russia\\}

\date{Accepted 2023 December 14. Received 2023 November 20; in original form 2023 September 28}

\pubyear{2023}

\begin{document}
\label{firstpage}
\pagerange{\pageref{firstpage}--\pageref{lastpage}}
\maketitle

\begin{abstract}
We present the results of the spectroscopic study of a chemically peculiar star HD~152564. Using medium-resolution ($R=37000$) observations obtained with the HRS spectrograph mounted on the South African Large Telescope we determined atmospheric parameters \te=11950$\pm200$~K and \logg=3.6$\pm0.2$~dex. Abundance analysis revealed mild deficiency of the light elements and an overabundance of up to $\sim$2 dex of metals with greatest excess for the silicon. With these characteristics HD~152564 is a typical member of the silicon subgroup of Ap stars. Rotational modulation of the light curve and line profiles of HD~152564 are typical for inhomogeneous surface distribution of elements in its atmosphere. We performed multi-element Doppler imaging of the HD~152564 surface. Abundance maps constructed for \ion{He}{}, \ion{O}{}, \ion{Mg}{}{}, \ion{Si}{}, and \ion{Fe}{} revealed concentration of these elements in a sequence of equatorial spots as well as in the circumpolar rings. The photometric maximum of the light curve coincided with the visibility of two most overabundant silicon spots. Abundances determined from the different ionisation stages of \ion{Fe}{} and \ion{Si}{} show clear evidence for vertical stratification of these elements in HD~152564 atmosphere. Meanwhile, the horizontal distribution of silicon reconstructed from the lines of different ionisation stages and excitation energies appeared to be identical with increasing average abundance deeper in atmosphere.
\end{abstract}




\section{Introduction}

One of the distinctive characteristics of the magnetic Ap/Bp stars among other chemically peculiar (CP) stars is their low amplitude photometric and spectroscopic variability caused by the inhomogeneous (spotted) distribution of chemical elements over their surfaces. Abundances of the iron peak and heavier elements in spots on magnetic Ap/Bp stars can exceed the solar ones by several orders of magnitude. The modified chemical composition affects the local thermal structure of the atmosphere inside the spot. Rotation of the spotted star therefore leads to modulation of both the spectral line profiles and the emergent flux. This is how the variability of Ap/Bp stars was initially explained in the framework of the oblique rotator model \citep{Babcock_1949,Stibbs_1950}.

The origin of the peculiar surface composition in Ap/Bp stars was explained within the framework of selective atomic diffusion mechanism \citep{Michaud_1970}. In a stable non-convective atmosphere selective diffusion leads to formation of the vertical abundances gradients: radiatively supported elements are pushed closer to the stellar surface, while the rest sink under the gravitational force. At the same time, in the presence of magnetic field, the suppressed transversal migration of ions leads to the formation of inhomogeneous horizontal distribution or chemical spots \citep{Michaud_2015}. Consequently, the distribution of elements in the atmospheres of Ap/Bp stars has a complex three-dimensional character. 

In chemically inhomogeneous atmosphere effects of the locally increased continuous opacity, line blanketing and departures from the local thermodynamic equilibrium (LTE) consequently alter the thermal structure and pressure distribution. Long ago \citet{Strom_1969} has shown that $\sim$1.5 dex increase in the silicon abundance in the line formation region of a star with \te=10000 K leads to a change in the thermal structure of the atmosphere ("overheating" in the upper layers), corresponding to an increase in the effective temperature by $\sim$1000 K, and to a significant flux redistribution in the ultraviolet region. The modern analysis by \citet{Khan_2007} employing the most sophisticated line-by-line (LL-) opacity treatment generally confirmed the significant influence of the \ion{Si}{}, \ion{Fe}{} and \ion{Cr}{} abundances on the thermal structure of the atmospheres of CP stars.

Hence, accurate modelling of the radiation of Ap/Bp stars requires calculations of the atmospheric structure, accounting for the effects of inhomogeneous chemical composition. Information about the horizontal distribution of elements can be obtained with a Doppler Imaging (DI) technique, 
while the vertical stratification profiles can be constructed by modelling the selective diffusion.

DI is a rigorous and powerful numerical technique based on inversion of the line profile variations, which allows to reconstruct the abundance distribution over the surface of a rotating star \citep[see][for review]{Piskunov_2008,Kochukhov_2016}. Using abundance distribution maps reconstructed with DI and chemically-homogeneous atmospheric models, the emergent flux was modelled for some Ap stars, and the crucial contribution of silicon and iron spots to the observed photometric variability was confirmed \citep[e.g][]{Krivoseina_1980,Krtichka_2007,Krtichka_2009,Krtichka_2012}. At the same time, accounting for the vertical elemental gradients leads to the successful reproduction of the peculiar line profiles and line strengths in magnetic Ap/Bp stars \citep{Ryabchikova_2003}. The next step for even more realistic modelling of the observed properties of the Ap/Bp stars should be a self-consistent modelling of both horizontal and vertical abundance distributions.

HD~152564 (=MX TrA) is a bright but little studied southern Ap star. It was classified as ApSi star in the Michigan Spectral Survey \citep{Houk_1975}. The star is recorded in the Washington Double Star Catalog \citep{Worley_1997} with an optical satellite at 1\arcsec~ separation and $\sim$3 mag. fainter. HD~152564 was also suspected as an SB1 system based on a few measurements that showed $\sim$20~\kms~ change of radial velocity over three years \citep{Levato_1996}. However, the low-amplitude photometric variability discovered in Hipparcos data was clearly consistent with the 2.163$^d$ rotational period, and hence HD~152564 was classified as $\alpha^2$ CVn-type variable \citep{Paunzen_1998}. The stellar parameters of HD~152564 presented in literature \citep{Netopil_2017}: \te=12500~K, \logg=3.76, were determined from the Str\"{o}mgren-Crawford and Geneva photometry and it is generally consistent with the assigned MK spectral class. Determinations of the projected rotational velocity \vsini\ gave somewhat controversial results between 75 and 103~\kms~ \citep{Levato_1996,Netopil_2017}. However, the star was not previously studied using high-resolution spectroscopy, and no attempts of magnetic measurements with spectropolarimetry were made.

Our observations from 2019 show that the star exhibits moderate rotational broadening and significant line profile variability, modulated with the rotational phase. HD~152564 also demonstrates a clear evidence for vertical abundance gradients. Hence, it is a promising target for investigating the effects of inhomogeneous three-dimensional abundance distribution on the emergent radiation properties. In this paper we present first results of the analysis including the atmospheric parameters and the chemical composition determination as well as the surface chemical maps obtained with the DI technique.

\section{Observations and data reduction}

\begin{table}
\caption{Log of the SALT spectroscopic observations of HD 152564}
\label{tab1}
\begin{tabular}{lccc}
\hline
Date & JD, 2450000+ & Exposure, s & Phase$^*$ \\
\hline
\hline
2019.04.17 &  8591.472  & 60 &  0.979 \\
2019.04.18 &  8592.468  & 60 &  0.439 \\
2021.03.24 &  9298.652  & 4$\times$60 &  0.795 \\
2021.03.25 &  9299.539  & 4$\times$60 &  0.205 \\
2021.03.29 &  9303.570  & 4$\times$60 &  0.067 \\
2021.03.30 &  9304.525  & 4$\times$60 &  0.509 \\
2021.04.03 &  9308.518  & 4$\times$60 &  0.354 \\
2021.04.08 &  9313.586  & 4$\times$60 &  0.696 \\
2021.04.09 &  9314.596  & 4$\times$60 &  0.163 \\
2021.04.30 &  9335.621  & 4$\times$60 &  0.880 \\
\hline
\end{tabular}
\bigskip

\emph{*} Ephemeris $JD(max.light)=2458647.7774+2^d.1639\cdot E$
\end{table}    

Spectral observations of HD~152564 were carried out with the use of the
High Resolution Fibre \'echelle  Spectrograph
\citep[HRS;][]{2008SPIE.7014E..0KB,2010SPIE.7735E..4FB,2012SPIE.8446E..0AB,2014SPIE.9147E..6TC}
at South African Large Telescope \citep[SALT;][]{2006SPIE.6267E..0ZB,2006MNRAS.372..151O}.
The HRS is a thermostable double-beam \'echelle spectrograph, with the entire optical part housed in a vacuum chamber
to reduce the impact of the thermal and mechanical variations.
The blue arm of the instrument covers the spectral range of 3735--5580~\AA,
and the red arm covers the spectral range of 5415--8870~\AA, respectively.
The spectrograph can be used in low (LR, R$\approx$14,000--15,000),
medium (MR, R$\approx$36,500--39,000), and high (HR, R$\approx$67,000--74,000) resolution modes
and it is fed with two fibers (object and sky fibers).
During observations of HD~152564 the MR mode was used with 2\farcs23 angular diameter of the fibers.
The CCD detectors in the blue and the red arms were used without binning.
The journal of observations is presented in Table~\ref{tab1}. 

The primary HRS data reduction was performed automatically using
SALT standard pipeline \citep{2010SPIE.7737E..25C} 
and following \'echelle data reduction was done using \textsc{MIDAS}-based HRS pipeline 
described in details in \citet{2019AstBu..74..208K}. The final normalisation to continuum level and merging of \'echelle orders was made with custom software. The wavelength scale was transformed to the laboratory rest-frame by applying the barycentric correction and correction of the radial velocity of the star, measured using hydrogen lines in the stellar spectrum.

\section{Results}

\subsection{Atmospheric parameters}\label{sect:atm_param}

The atmospheric parameters of HD~152564 were determined iteratively based on the simultaneous fitting of the Balmer line profiles and spectral energy distribution (\SED). The hydrogen line profiles in the considered temperature range are sensitive to both the effective temperature, \te\ and the surface gravity, \logg. Therefore, fitting of the \SED, which was constructed using archival photometry (see below) helps breaking degeneracy of the solution. The magnitude of the Balmer jump and the slope of the Paschen continuum measured from the SED were used as independent constraints for \logg\ and \te. In addition, the \SED\ fitting made it possible to determine the stellar radius and to refine the value of interstellar extinction.

The initial parameters in our fitting: \te$\approx$13500~K and \logg$\approx$4.0 were estimated from the Str\"{o}mgren-Crawford \citep{Eggen_1977} and Geneva \citep{Paunzen_2021} photometry of
HD~152564 with \textsc{TempLogG} program \citep{Kaiser_2006}. The \citet{Moon_1985,Napiwotzki_1993,Balona_1994} calibrations were used for the Str\"{o}mgren system and \citet{Kunzli_1997} for the Geneva ones. Then, the effective temperature was reduced to \te$=$12500~K \citep{Netopil_2017} as a result of the special correction for peculiar stars. Reduction of the effective temperature is caused by the flux redistribution in the chemically-enriched atmosphere. However, an accurate account for such effects and hence the robust parameter determination for CP stars is only possible by means of the spectral synthesis.

The LTE spectrum synthesis of the three Balmer lines $H\beta$, $H\gamma$, $H\delta$ was performed with the \textsc{Synth3} code \citep{Kochukhov_2007} using the atomic data from the \vald\ database \citep{Piskunov_1995,Ryabchikova_2015,Pakhomov_2019}. The $H\alpha$ line was excluded from analysis because of strong possible NLTE effects. The plane-parallel atmospheric model employed in spectral synthesis was calculated with the \textsc{LLmodel} code \citep{Shulyak_2004}, which includes individual lines in the total opacity calculations. Such a line-by-line treatment takes into account the opacity distribution in the atmospheres of chemically peculiar stars with the abundances anomalies in the line forming regions. At each iteration the atmospheric models were recomputed taking into account the current abundance estimate for the major contributors to the opacity: \ion{H}, \ion{He}, \ion{Si}, \ion{Fe}, \ion{Cr}. The aim of the iterative procedure was to determine the set of parameters, which provided the best match between the synthetic spectrum and the observed outer wings of the hydrogen lines avoiding their NLTE cores.

Simultaneously to line profiles fitting, the theoretical emergent flux was calculated on each iteration and compared with the observed SED of HD~152564. Observational SED was constructed in the $0.15-22~\mu m$ range using the ultraviolet spectrophotometry from the TD1 \citep{Thompson_1978} catalogue, infrared data from the Two Micron All Sky Survey (2MASS, \citep{Skrutskie_2006}) and Wide-field Infrared Survey Explorer (WISE, \citep{Wright_2010}) catalogues. The optical $UBVRI$ photometry was compiled from the \textsc{Vizier-SED} service. According to \citet{Lallement_2019} dust extinction map, reddening is almost negligible on the line of sight toward HD~152564. At the stellar distance of $D=191.5^{+9}_{-8}$ pc obtained from the inversion of the \textsc{Gaia DR3} parallax \citep{Gaia_2021}, the map provides a reddening $E(B-V)=0.027\pm0.15$. This reddening can be converted to the visual extinction $A_V=0.084\pm0.065$ using the standard total-to-selective absorption ratio of $R_V=3.1$. The observational photometric data were dereddened using \citet{Fitzpatrick_2019} Galactic extinction curve, adopting the obtained $A_V$ value. Theoretical fluxes were scaled to the given distance $D$, while the stellar radius $R_*$ was fitted to achieve a best match between the observed and the theoretical luminosities.

Atmospheric abundances for model calculations were derived from the spectral synthesis of the HD~152564 spectra, averaged over rotational phase (Section~\ref{sect:lte_abund}).

\begin{figure}
	\includegraphics[width=1.0\columnwidth, clip]{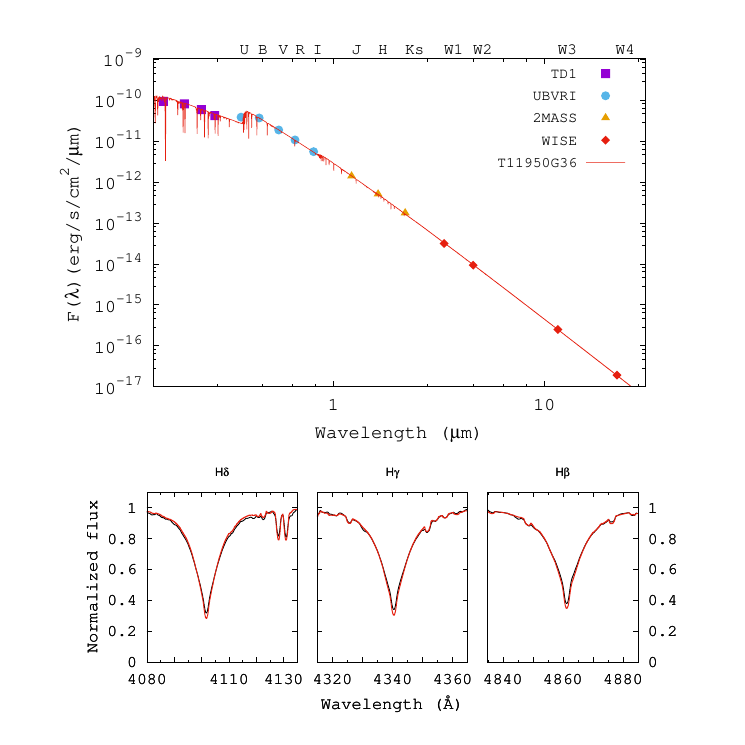}
    \caption{The SED (upper panel) and the Balmer lines (lower panel) in HD~152564. Observational data are shown by points in the panel with SED and by black curve in the panel with the line profiles. The best-fit synthetic spectrum is shown as red curve.}
    \label{fig:1}
\end{figure}

The iterative procedure converged to the following parameters: \te=11950$\pm$200K, \logg=3.6$\pm0.2$ and $R/R_{\odot}=3.8$. Spectral synthesis with these parameters results in an excellent representation of the hydrogen line profiles and of the shape of the observed SED (Fig.~\ref{fig:1}). With the spectroscopically determined \te\ and $R_*$ we have calculated the luminosity of the star $\log (L/L_{\odot})=2.42$ from the Stefan-Boltzmann law. From the \logg\ and $R_*$ we derived its mass as $M/M_{\odot}=2.1$. In order to determine the projected rotational velocity \vsini~ and macroturbulence \vmacro~ we used numerous metallic absorption lines in few selected spectral windows and fitted their broadening with \textsc{BinMag6} \citep{Kochukhov_2018c} tool. We found \vsini=69$\pm$2~\kms\ and zero macroturbulence. The final set of parameters of HD~152564 is summarised in Table~\ref{tab2}.

The classical procedure for determination of the microturbulence \vmicro~is based on minimising the dependence of the lines equivalent width on the individual abundances for a given element. This procedure is not applicable for a magnetic Ap star due to the possible differential magnetic intensification of the lines. However, the analysis of the weakly-magnetic peculiar stars usually gives zero microturbulence \citep[e.g][]{Kochukhov_2006,Potravnov_2023}, which is consistent with the theoretical expectations. Therefore, for HD~152564 we also adopt \vmicro=0 \kms. There is no information about the magnetic field of the star and given its rotational velocity and line blending there is no chance to obtain a reliable estimate of the field strength from unpolarised (intensity) spectra. We also performed some numerical tests introducing non-zero microturbulence to evaluate the possible effect of the putative magnetic field and its effect on the Doppler maps (see Sect. \ref{sect:dopp_im}).

\begin{table}
\caption{Parameters of HD~152564}
\label{tab2}
\begin{tabular}{lc}
\hline
Parameter & Value \\
\hline
\hline
\te & 11950$\pm$200 K \\
\logg  & 3.6$\pm$0.2 dex \\
\vmicro  & 0.0~\kms \\
\vmacro  & 0.0$\pm$1.5~\kms \\
\vsini    & 69$\pm$2~\kms \\
$\log (L/L_{\odot})$ & 2.42 \\
$M/M_{\odot}$       & 2.1 \\
$R/R_{\odot}$       & 3.8\\
\hline
\end{tabular}
\end{table}    

\subsection{Abundance analysis}\label{sect:lte_abund}

\begin{figure}
	\centering
	\includegraphics[width=1.0\columnwidth]{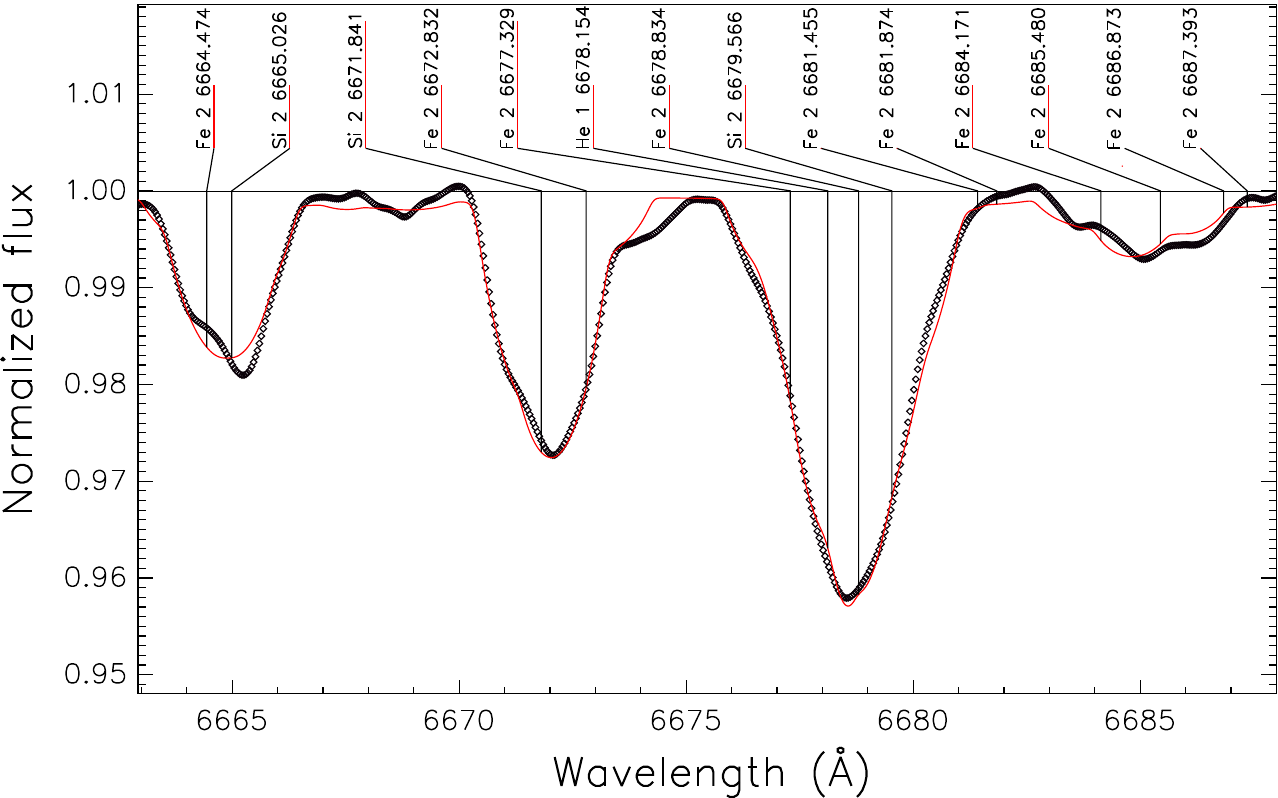}
    \caption{Example of the fitting procedure in spectral region with \ion{Si}{II}, \ion{He}{I} and \ion{Fe}{II} lines. Observed averaged spectrum 
				is represented by open symbols, synthetic spectrum is shown by a red curve.}
    \label{fig:2}
\end{figure}

\begin{figure}
	\centering
	\includegraphics[width=1.0\columnwidth]{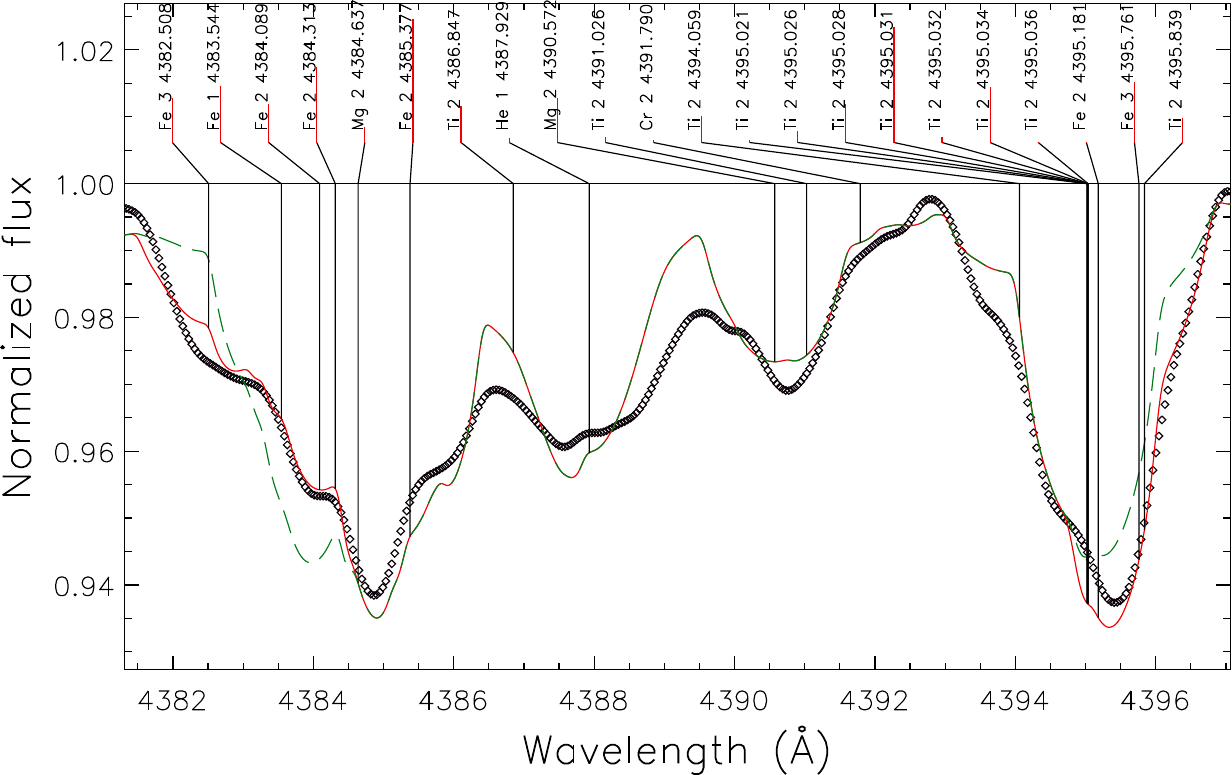}
    \caption{Portion of the HD~152564 spectrum containing \ion{Fe}{} lines from three ionisation stages. Observed averaged spectrum is shown by open symbols. Spectral synthesis of this blend calculated with the iron abundance deduced from the \ion{Fe}{II} lines, is shown by green dashed curve. Red curve represents synthetic spectrum calculated for individual abundances for different \ion{Fe} ionization stages from Tab.~\ref{tab3}.}
    \label{fig:3}
\end{figure}

Atmospheric abundances in HD~152564 were derived by fitting small spectral windows dominated by one of a few unblended or mildly blended lines. The list of these lines together with the atomic parameters and the corresponding references as well as the resulting abundances are given in Table~\ref{lines}, which is available in its entirety in a machine-readable form in the online version. 
Individual abundances were deduced from the fitting the synthetic line profiles to the observed ones in the mean spectrum averaged over all rotational phases employing \textsc{BinMag6} tool under LTE assumption. We performed NLTE calculations only for oxygen because of the expected large NLTE effects for \ion{O}{I} infrared triplet. Oxygen NLTE analysis was done with an updated model atom from  \citet{2000A&A...359.1085P} and \citet{sitnova_o}.
Table~\ref{tab3} summarises the mean abundances for 9 elements with the errors estimated as a standard deviation of individual measurement from the mean. All abundances are presented in logarithmic scale  $\log (A)_X = \log(N_{X}/N_{tot})$, where $\log N_{tot}$=12.04 for solar helium abundance. The last column of Table~\ref{tab3} contains the most recent data on solar photospheric abundances \citep{2021A&A...653A.141A}. An example of the fitting procedure is shown in Figs.~\ref{fig:2}-~\ref{fig:3}.

In the HD~152564 spectrum silicon lines represent two ionisation stages \ion{Si}{II-III}, while for iron the lines from three ionization stages \ion{Fe}{I-III} can be detected (Fig.~\ref{fig:3}). However, \ion{Fe}{I} lines are weak and appear in complex blends only. For both \ion{Si}{} and \ion{Fe}{} we notice a significant increase of abundances with ionisation stage (see Tab.~\ref{tab3}). For example, \ion{Si}{II} lines yield $\log (A)_{Si}$=-3.62 while \ion{Si}{III} lines are consistent with $\sim$1.1 dex larger abundance. The situation is even more prominent in case of \ion{Fe}{} lines with $\sim$1.5 dex difference between abundances derived from \ion{Fe}{I} and \ion{Fe}{III} lines. In order to estimate the possible influence of NLTE effects on derived abundances, we performed a set of preliminary calculations using model atoms and methods described in \citet{Mashonkina_2020} for silicon and in \citet{Mashonkina_2011,Sitnova_2018} for iron. We found that NLTE corrections can decrease the abundance difference between \ion{Si}{II/III} by $\sim0.3-0.5$ dex and by $\lesssim0.15$ dex for \ion{Fe}{I/II}. Hence, the difference in silicon and iron abundances deduced from the different ionisation stages cannot be fully attributed to the NLTE effects. The NLTE corrections partly reduces the discrepancy and  we interpret the remaining differences as a signature of vertical abundance stratification in HD~152564 atmosphere.

\begin{table}
\caption{Average LTE abundances in HD~152564}
\label{tab3}
\tabcolsep=12pt
\begin{tabular}{lccc}
\hline
Ion & $N_{lines}$ & $\log (A)_X$ & $\log (A)_X^{\odot}$ \\
\hline
\hline
\ion{He}{I} & 6 & -1.61$\pm0.09$ & -1.13 \\  
\ion{C}{II} & 3 & -2.70$\pm0.04$ & -3.58 \\  
\ion{O}{I}  & 2 & -4.03$\pm0.14$* & -3.35 \\  
\ion{Mg}{II} & 4 & -4.97$\pm0.08$ & -4.49 \\
\ion{Al}{II} & 2 & -6.72$\pm0.20$ & -5.61 \\
\ion{Si}{II} & 11 & -3.62$\pm0.22$ & -4.53 \\
\ion{Si}{III} & 3 & -2.55$\pm0.13$ & -4.53 \\  
\ion{Ti}{II} & 6 & -6.13$\pm0.14$ & -7.07 \\ 
\ion{Cr}{II} & 7 & -5.50$\pm0.16$ & -6.42 \\ 
\ion{Fe}{I} & 4 & -4.50$\pm0.25$ & -4.58 \\  
\ion{Fe}{II} & 12 & -3.85$\pm0.12$ & -4.58 \\
\ion{Fe}{III} & 2 & -2.98$\pm0.30$ & -4.58 \\
\hline
\end{tabular}

{\it Note.} * - NLTE abundance is given for  oxygen.
\end{table}

\subsection{Photometric variability}\label{sect:phot_var}

Photometric variability of HD 152564 was investigated using the \textit{TESS} spacecraft observations obtained during its Cycle 1 (Sector 012). The 120-s cadence light curves, processed with the Science Processing Operations Center (SPOC) pipeline, were retrieved from the Mikulski Archive for space telescopes (MAST)\footnote{https://mast.stsci.edu/portal/Mashup/Clients/Mast/Portal.html}. Because the effective wavelength of the \textit{TESS} broadband filter nearly coincides with those of Cousins $I_C$ filter we used the formula $I_{mag}=-2.5\log(SAP\_FLUX)+20.44$ for converting the differential fluxes to magnitudes in Vega system. However, due to difficulties of the absolute photometric calibration of \textit{TESS} data, we caution against direct comparison of these magnitudes with other photometric data. The HD~152564 light curve shows clear periodic light changes with $\Delta I\approx0^m.03$ amplitude. Our frequency analysis revealed single powerful peak on the Lomb-Scargle periodogram corresponding to the period $P=2^d.1639$, which is in excellent agreement with those previously found by \citet{Paunzen_1998}. The ephemeris for the maximum light is: $JD(max.light)=2458647.7774+2^d.1639\cdot E$. The phase curve folded with this period shows clear quasi-sinusoidal shape typical for other Ap/Bp stars with $\alpha^2$ CVn-type variability and implies its rotational origin. Assuming the rigid rotation of the star and combining spectroscopically determined \vsini=69~\kms~with the determined period we found the inclination $i=51^\circ$ for the stellar rotational axis.

\begin{figure}
	\includegraphics[width=1.0\linewidth]{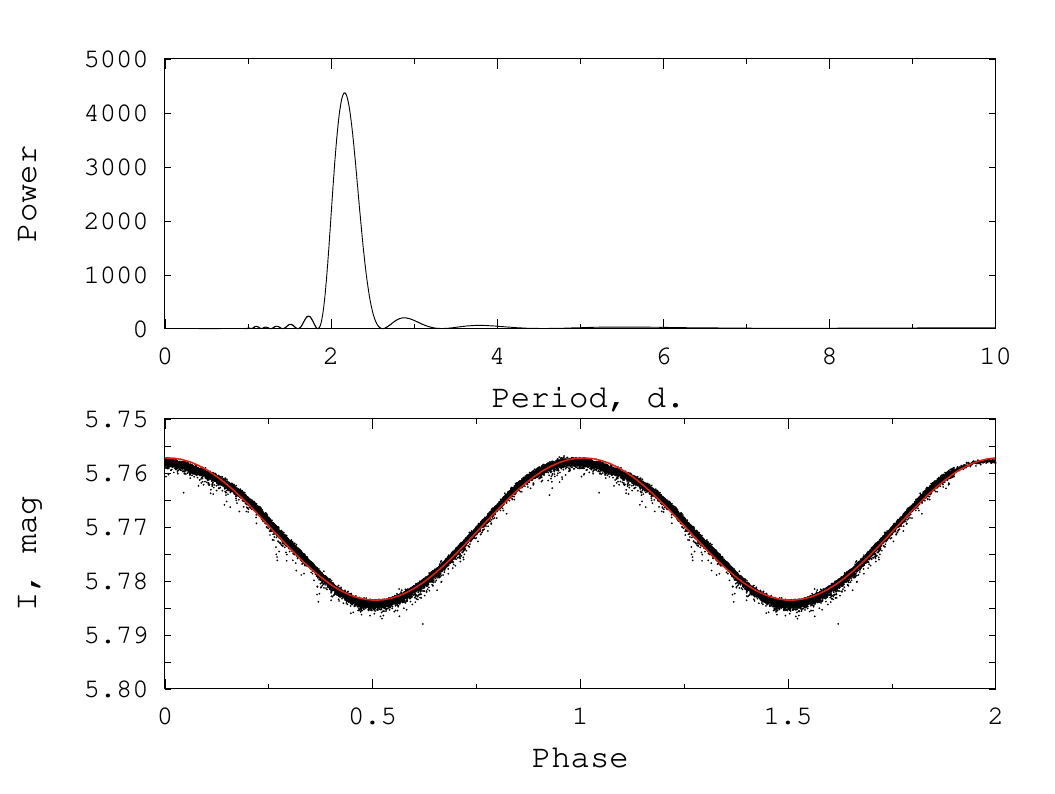}
    \caption{Power spectrum (upper panel) of HD 152564 photometric variability and phase curve (lower panel) folded with $P=2^d.1639$ period. The sinusoidal fit of the phase curve is shown by red solid curve.}
    \label{fig:4}
\end{figure}

\subsection{Doppler imaging}\label{sect:dopp_im}
The surface abundance distribution of five elements: \ion{Si}{}, \ion{He}{}, \ion{Fe}{}, \ion{Mg}{}, \ion{O}{}, which exhibit prominent line variability in the HD~152564 spectrum and/or serve as important markers for selective atomic diffusion was studied using DI technique. We used the \textsc{INVERS11} code developed by N. Piskunov, which solves the line-profile inversion problem based on theoretical treatment generally described in \citet{Piskunov_1993}. This code is a non-magnetic version of the \textsc{INVERS10} program, which is described in detail in \citet{Piskunov_2002}. \textsc{INVERS11} solves the radiative transfer to compute specific intensities for each spectral line within the selected wavelength ranges. This allows to use spectral line blends and to perform the multi-element fitting. Calculations are performed on each iteration using the spherical grid covering the visible stellar hemisphere.

The input data for the DI inversion include the atmospheric model, \vsini, inclination of the rotational axis $i$, atomic parameters for the lines to be fitted and an initial guess for the surface distribution of the mapped elements. 

For the spectral synthesis in the \textsc{INVERS11} we used the \textsc{LLmodel} model, calculated with the final set of atmospheric parameters and with the average chemical composition of HD~152564 (Tab. \ref{tab2} and Tab. \ref{tab3}). Projected rotational velocity \vsini=69~\kms\ and inclination angle $i=51^{\circ}$ were also determined in our analysis. Atomic data for the examined spectral lines were obtained from the \vald\ database. We did imaging of single element distribution whenever we could identify unblended lines and multi-element mapping when we could only use blended lines. For all elements but \ion{He}{} we used few lines in order to increase the validity of the obtained map. For helium we employed single unblended \ion{He}{I} 5876\AA\, line. The atomic data for the transitions used in DI are summarised in Tab. \ref{tab4}. We started with the two sets of initial abundances of a given element: solar ones and those from Table~\ref{tab3} and obtained nearly identical abundance maps. For silicon, which demonstrated the difference of $\sim1.0$ dex in abundances determined from the lines of different excitation energies and ionisation stages, we considered separately these moderate- and high-excitation lines with different initial abundances adopted.

It is possible to estimate the minimal size of the grid for fitting local intensities without the loss of the spatial information. The angular size of the resolution element at stellar equator was estimated using the standard approach based on comparison of the width of the rotationally-broadened profile with the  instrumental profile of our observations \citep[e.g.][]{Kochukhov_2018b}:

\begin{equation}
 \delta l=90^{\circ}\frac{c}{R\cdot V\sin(i)}
\end{equation}

Substituting the resolving power $R=37000$ of HRS spectrograms, projected rotational velocity \vsini=69 \kms~ and speed of light $c$ we obtained $\delta l=10.6^{\circ}$ which corresponds to $\approx$36 resolution elements on the equator. In a spherical grid, this corresponds to a minimum number of elements $N_{grid}=440$. Actually we performed calculations over three times denser 1436-elements grid covering the whole star, in order not to lose the information at high-latitude regions, where spots produce a smaller relative distortion of the Doppler profile. The choice of the optimal value of the regularisation parameter in the DI inversion procedure was made in the same way as described by \citet{Kochukhov_2017}. Fitting the spectroscopic time-series for 10 equidistant rotational phases of HD~152564 allowed to reconstruct the abundance maps for each element, as detailed below.

In order to check the influence of the possible magnetic field of HD~152564 on the abundance maps we performed numerical tests introducing the pseudo-microturbulence. It was shown that for the fields $B\lesssim2-4$ kG magnetic intensification can be replaced with reasonable accuracy by the parameter of pseudo-microturbulence, expressed as $v_{magn}=1.4\times10^{-7}\lambda\mid{B}\mid g_{eff}$ \citep{1996A&A...308..886K}. Here $B$ is the modulus of magnetic field, $g_{eff}$ - effective Lande factor. We performed calculations for silicon and iron lines adopting the magnetic field strength $B\lesssim2$ kG and Lande factors retrieved from the \vald. Typically the value of pseudo-microturbulence did not exceed $\sim2.2$ \kms. Comparison with the maps calculated assuming \vmicro=0`\kms\ showed their general geometrical consistency and difference only in abundance scale, as was expected. Hence, the presence of an unaccounted uniform magnetic field of a few kG strength and possible errors in microturbulent velocity should not affect our results on the surface abundance distribution in HD~152564.

The results of DI are discussed in Sect.~\ref{sub:Si}-\ref{sub:O} and presented in Figures~\ref{fig:5}-\ref{fig:10} in both Mercator and spherical projections (for 4 phases). The coordinate system is set so that zero longitude on the Mercator map corresponds to the central meridian in the spherical plot for phase $\varphi=0$, the phases are counted according to the ephemeris in Sect.~\ref{sect:phot_var}. It should be noted, that with the derived inclination of stellar rotational axis $i=51^{\circ}$ the region of the stellar surface south of latitude $\phi=-40^\circ$ remains invisible and the results of the Doppler inversion in this zone on the Mercator map should be ignored. In the figures this boundary is marked by a dashed line. 

\begin{table}
\caption{List of elements and lines used for DI}
\label{tab4}
\tabcolsep=3pt
\begin{tabular}{lccccc}
\hline
Element & Line, \AA\, & $E_{low}$, eV & $\log gf$ & $\log \Gamma_4$ & Comment \\
\hline
\hline
\ion{Si}{II} & 5041.024 & 10.07 & 0.030 & -4.80 & bl. \ion{Fe}{II}, \ion{Ni}{II} \\
             & 6347.109 & ~8.12 & 0.150 & -5.08 & bl. \ion{Fe}{II}, \ion{Mg}{II}\ \\ 
             & 6660.532 & 14.50 & 0.162 & -5.92 &  \\ 
             & 6665.026 & 14.49 &-0.240 & -5.54 & w.bl. \ion{Fe}{II}  \\
             & 6671.841 & 14.53 & 0.410 & -5.58 &   \\
\ion{Si}{III}& 4552.622 & 19.02 & 0.290 &  0.00 &  \\
\\
\ion{He}{I} & 5875.599 & 20.96 & -1.511 & -4.72 &  \\
& 5875.614 & 20.96 & -0.341 & -4.72 &  \\
& 5875.615 & 20.96 & 0.409 & -4.72 &  \\
& 5875.625 & 20.96 & -0.341 & -4.72 &  \\
& 5875.640 & 20.96 & 0.139 & -4.72 &  \\
& 5875.966 & 20.96 & -0.211 & -4.72 & w.bl.atm. $H_2O$ \\

\\
\ion{Fe}{II} & 4595.673 & 2.85 & -4.583 & -6.53 &  \\
 & 4596.009 & 6.22 & -1.956 & -6.53 &  \\
 & 4598.485 & 7.80 & -1.536 & -5.82 &  \\
 & 4731.447 & 2.89 & -3.127 & -6.53 &  \\

\\
\ion{Mg}{II} & 4481.126 & 8.86 & 0.749 & -4.70 & bl. \ion{Fe}{I-II}  \\
 & 4481.150 & 8.86 & -0.560 & -4.70 &  \\
 & 4481.325 & 8.86 & 0.590 & -4.70 &  \\
\\
\ion{O}{I} & 7771.944 & 9.15 & 0.369 & -5.50 &  \\
 & 7774.166 & 9.15 & 0.223 & -5.50 &  \\
 & 7775.388 & 9.15 & 0.002 & -5.50 &  \\

\hline
\end{tabular}
\end{table}

\subsubsection{Silicon}\label{sub:Si}
Due to significant silicon mean overabundance in HD~152564 atmosphere, its line spectrum is represented by strong lines of both moderate and high excitation. For example \ion{Si}{II} \eu{4s}{4}{P}{\circ}{} -- \eu{4p}{4}{D}{}{} and \ion{Si}{III} \eu{4s}{3}{S}{}{} -- \eu{4p}{3}{P}{\circ}{} transitions with the lower 
levels at 14.5 and 19 eV respectively, which are normally weak in the given temperature domain, are well detectable in the spectrum of HD~152564 and suitable for analysis. A detailed silicon linelist used for our DI calculations is 
given in Table \ref{tab4}. The inversion procedure was performed separately for \ion{Si}{II} 5041~\AA\,, \ion{Si}{II} 6347~\AA\,, \ion{Si}{III} 4552~\AA\, lines and simultaneously for the lines of \ion{Si}{II} 6660, 6665, 6671~\AA\, triplet. In the latter case we artificially adjusted the oscillator strengths $\log gf$ to equalise the abundances deduced from these transitions to the average value and thus compensate for the possible errors in atomic data.

\begin{figure*}
	\includegraphics[width=1.0\linewidth]{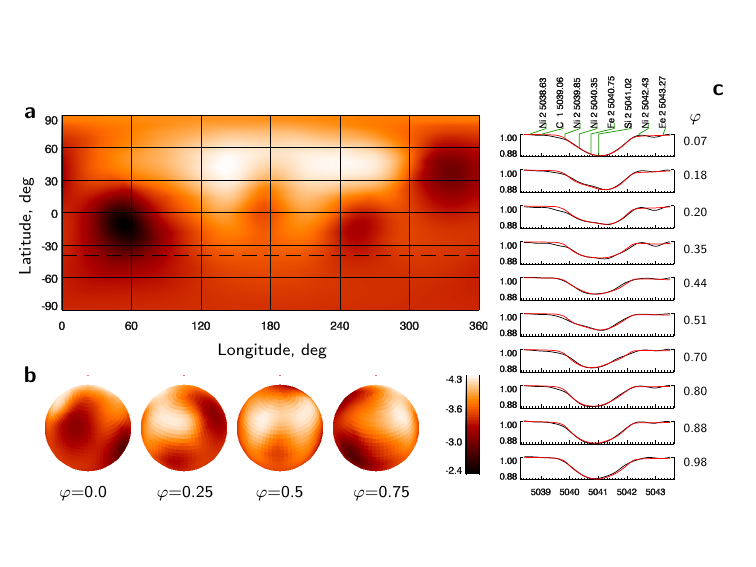}
    \caption{\ion{Si}{} distribution in HD~152564 plotted in Mercator (panel \textit{a}) and spherical (panel \textit{b}) projections, obtained from DI of \ion{Si}{II} 5041~\AA\, line. The abundance scale is encoded in the colorbar to the right from the spherical plot. Line profile fitting is shown in panel \textit{c}, phases are labelled at the right side of the plot.}
    \label{fig:5}
\end{figure*}

Abundance maps deduced from the moderate excitation \ion{Si}{II} 5041~\AA\, (Fig. \ref{fig:5}) and 6347~\AA\ lines are quite similar to each other and reveal the concentration of the silicon in four spots extended along the equator. The most distinctive spots with silicon abundance reached $\log (A)_{Si}\approx-2.4$ dex are located between 30-80$^\circ$ and 310-350$^\circ$ in longitude. The former is shifted towards northern latitudes. Another two spots with somewhat lower abundance contrast are distributed between 150-270$^\circ$ longitude and form the horsehoe-like structure. The northern mid-latitudes are characterised by an extended strip with the near-solar silicon abundance. The \ion{Si}{} abundance derived from the moderate excitation lines and averaged over stellar surface is $\log (A)_{Si}\approx-3.3$ dex. 

\begin{figure*}
	\includegraphics[width=1.0\linewidth]{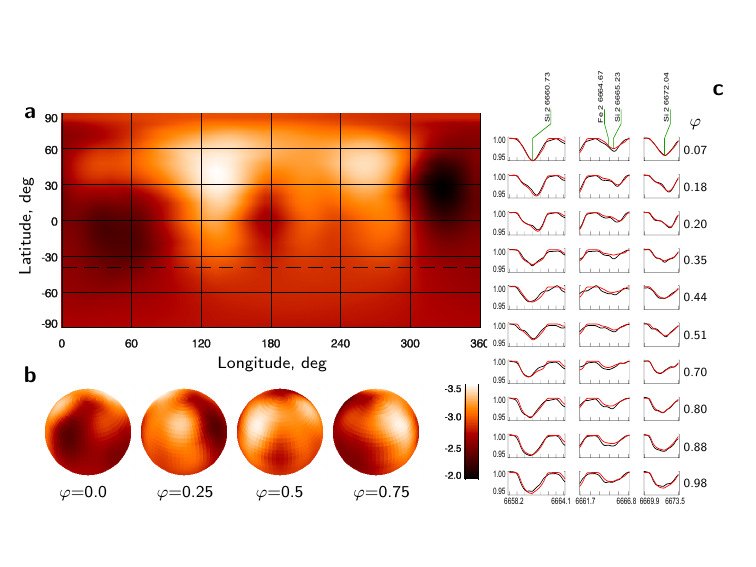}
    \caption{The same as in Figure \ref{fig:5} but for \ion{Si}{II} 6660, 6665, 6671~\AA\, lines.}
    \label{fig:6}
\end{figure*}

The \ion{Si}{} distribution over HD~152564 surface as it is traced by high excitation \ion{Si}{II} 6660, 6665, 6671~\AA\, (Fig.~\ref{fig:6} and \ion{Si}{III} 4552~\AA\, lines is essentially the same as the one inferred from the moderate excitation \ion{Si}{II} lines. Abundance maps obtained with the high excitation lines revealed the median abundance $\log (A)_{Si}\approx-2.7$ dex and four silicon patches with largest silicon abundance up to $\log (A)_{Si}=-2$ dex. The coincidence of the structures revealed on the abundance maps confirms the identical horizontal distribution of the silicon regardless of the depth in the atmosphere. However, the different abundances determined from the lines of different excitation clearly indicate vertical stratification of silicon.

\subsubsection{Helium}\label{sub:He}
\begin{figure*}
	\includegraphics[width=1.0\linewidth,clip]{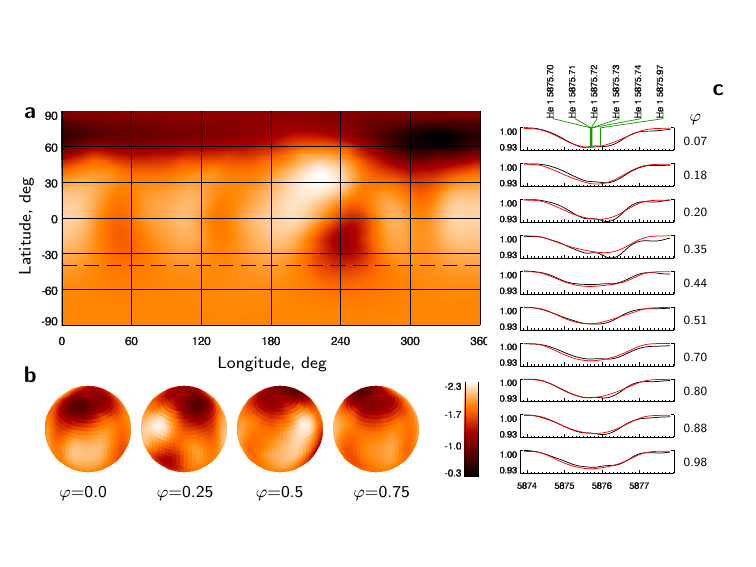}
    \caption{The same as in Figure \ref{fig:5} but for \ion{He}{I} 5876 ~\AA\, line.}
    \label{fig:7}
\end{figure*}

The helium distribution (Fig. \ref{fig:7}) in HD~152564 is characterised by general deficiency except for the polar region. Our maps show that regions of silicon overabundance along the equator are likely to coincide with the regions of helium depletion by up to $\log (A)_{He}\sim-2.3$ dex. However we should point that our \ion{He}{} map was constructed on the base of a single helium line and hence, it represents only preliminary view on the distribution of this element. The averaged helium abundance is sub-solar with $\log (A)_{He}\approx-1.6$ dex.

\subsubsection{Iron}\label{sub:Fe}
\begin{figure*}
	\includegraphics[width=1.0\linewidth]{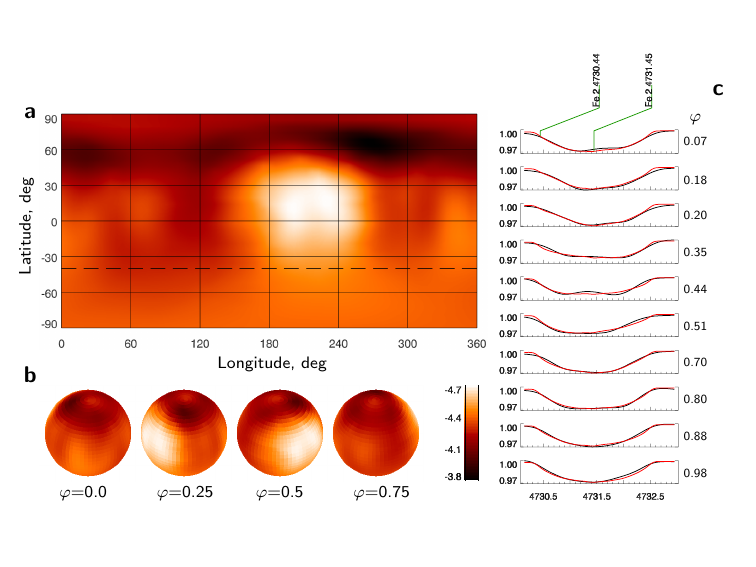}
    \caption{The same as in Figure \ref{fig:5} but for \ion{Fe}{II} 4731 ~\AA\, line.}
    \label{fig:8}
\end{figure*}

With INVERS11, we constructed iron distribution maps from a few unblended \ion{Fe}{II} lines of low to moderate excitation energies (see Tab. \ref{tab4}). We have also performed multi-element mapping including iron in blends (e.g. at 6347\AA\, with \ion{Si}{} as a dominant contributor). The resulting abundance maps are fairly close in abundance pattern. The iron abundance in HD~152564 varies over the surface in the $\sim-4.7$ to $-3.8$ dex range with strongest \ion{Fe}{II} overabundance in the circumpolar ring at the $\sim$70$^\circ$ latitude (Fig. \ref{fig:8}). Within this ring, the \ion{Fe}{} patch with the maximal concentration can be isolated in the 240-300$^\circ$ longitudinal range. The equatorial region (between 180-240$^\circ$ longitudes) generally possesses the near-solar or moderately depleted iron abundance.

\subsubsection{Magnesium}\label{sub:Mg}
\begin{figure*}
	\includegraphics[width=1.0\linewidth]{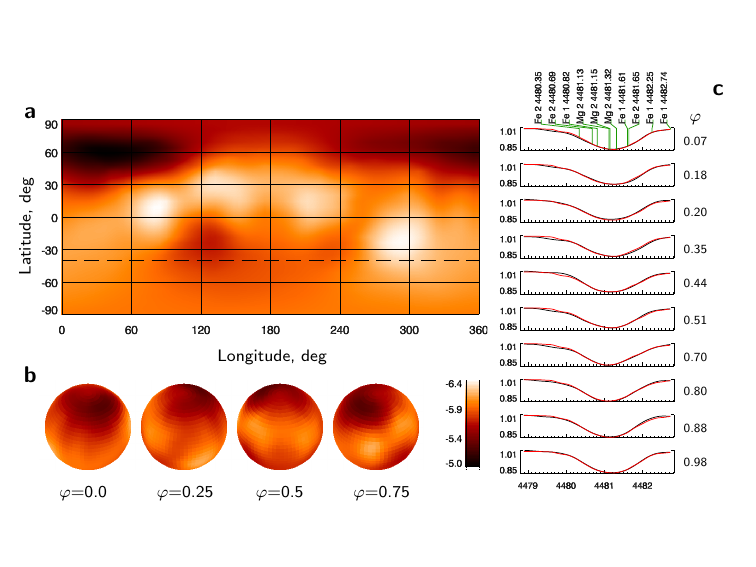}
    \caption{The same as in Figure \ref{fig:5} but for \ion{Mg}{II} 4481 ~\AA\, line.}
    \label{fig:9}
\end{figure*}

The magnesium distribution (Fig. \ref{fig:9}) deduced from fitting of the strong \ion{Mg}{II} line at 4481\AA\, and multi-element fitting of the 6347\AA\ blend demonstrates similar behaviour to \ion{Fe}{}. The \ion{Mg}{} abundance over stellar surface varies between -6.4 and -5.0 dex. The region of increased magnesium abundance (given its general depletion in HD~152564 compared to the Sun) coincides with the circumpolar ring of \ion{Fe}{} distribution at $\sim$70$^\circ$ latitude. The region of the highest abundance is, however, more blurred and somewhat shifted in longitude with respect to the iron patch. Areas of the greatest underabundance are located in the patchy structure along the equator.   

\subsubsection{Oxygen}\label{sub:O}
\begin{figure*}
	\includegraphics[width=1.0\linewidth]{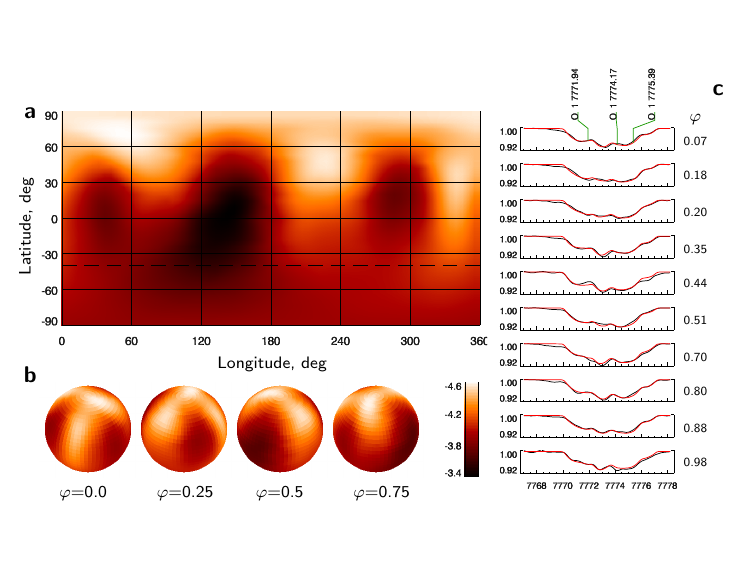}
    \caption{The same as in Figure \ref{fig:5} but for \ion{O}{I} 7774 ~\AA\, triplet.}
    \label{fig:10}
\end{figure*}

The fit of the partially-resolved \ion{O}{I} triplet at 7774\AA\ suggests that oxygen is concentrated in the three large spots in the equatorial region of HD~152564 between +30$^\circ$ and -30$^\circ$ latitudes and within the 30-50$^\circ$, 100-180$^\circ$ and 270-320$^\circ$ longitudinal ranges (Fig. \ref{fig:10}). Oxygen abundance at different phases ranges from -3.7 to -2.2 dex with a median value $\log (A)_{O}=-2.95$ dex. These values refer to LTE analysis. We tried to estimate the impact of NLTE corrections by changing the transition probabilities (TP) of the triplet lines based on the NLTE fitting of the averaged spectrum and then using these TPs in DI. The resulting map is almost the same in terms of geometry but with surface abundance scale decreased by $\sim$1 dex. This particular map and the line fits are shown in the Figure \ref{fig:9}. The distribution of oxygen resembles that of silicon, although it is difficult to claim an exact spatial coincidence of the regions with the overabundance of these elements.

\section{Discussion}

We determined atmospheric parameters and average LTE atmospheric abundances of HD~152564 using medium-resolution spectroscopy. The effective temperature is \te=11950 K and the prominent silicon overabundance up to $\sim$2~dex relative to the solar composition agrees well with the previous qualitative spectral classification of HD~152564. These parameters are very typical for the members of \ion{Si}{} subgroup of CP stars. Rotationally modulated photometric variability of HD~152564 as well as the line profile variations clearly visible in our spectroscopic time-series are suggestive for the inhomogeneous surface abundances distribution. 

Indeed, abundance maps inferred from the DI inversion of five elements show their strongly inhomogeneous distribution: contrasts up to few dex between large-scale overabundance patches and the rest stellar surface with near-solar (e.g. \ion{Si}{}, \ion{Fe}{}) or sub-solar (e.g. \ion{He}{}, \ion{Mg}{}) composition. Elemental distribution maps turn out to be different for different elements. In particular, silicon and oxygen are concentrated in spots located in broad latitudinal zones along the equator. However, the number of spots: 4 for silicon and 3 for oxygen, and their positions and sizes are not exactly the same. Helium, in contrast to silicon, reveals a depletion in the equatorial zone and one prominent high-latitude spot with a markedly increased abundance. We emphasize the anticorrelation of \ion{He}{} and \ion{Si}{} surface distributions, which is the well-known characteristic of magnetic Ap/Bp stars \citep[see, for example][]{1999A&A...348..924K,Briquet_2004,Oksala_2015}. Iron and magnesium show similar distributions, different from above mentioned elements. Both are concentrated in circumpolar rings with single high-contrast feature, although misaligned in longitude. The magnesium shows overall underabundance relative to the solar composition, while iron has a near-solar abundance over most of the surface with an overabundance in the polar region. The overall horizontal distribution of elements in HD~152564 is well within diversity observed in other mapped Ap/Bp stars. Structures and surface abundance contrasts revealed in our analysis of HD~152564 resemble those of HR 5264 (=HD~133880) performed with Zeeman-Doppler imaging, which also accounts for the magnetic field geometry \citep{Kochukhov_2017b}.

Comparison of the photometric variability of HD~152564 with the obtained longitudinal distribution of chemical spots reveals that phases of maximum light coincides with the appearance of the prominent silicon spots on the visible hemisphere of the star. This result is in qualitative agreement with the ones previously obtained by \citet{Krivoseina_1980} and \citet{Krtichka_2007,Krtichka_2009,Krtichka_2012}, who showed that silicon overabundance in spots produces local overheating of the upper atmosphere leading to the rotationally-modulated brightness changes. For HD~152564 the perspective goal is to model its flux variability using the reconstructed surface chemical maps and taking into account locally modified atmospheric structure.

HD~152564 shows pronounced evidence for vertical chemical stratification at least for silicon and iron - elements represented in spectrograms by lines of two (for \ion{Si}{}) or three (for \ion{Fe}{}) ionisation stages, which demonstrate an increase of abundance with depth in the atmosphere. This result is in agreement with other observational studies, which revealed a definite abundance difference deduced from \ion{Si}{II/III} lines  - the so-called "\ion{Si}{II/III}-anomaly"~ observed in spectra of early-A to late-B CP stars and sometimes in normal stars \citep{Bailey_2013}. In normal stars accounting for NLTE effect can completely remove this anomaly \citep{Mashonkina_2020}. In CP stars this anomaly together with \ion{Fe}{} abundance difference is usually considered as a manifestation of the vertical stratification, caused by selective atomic diffusion, although accounting for the NLTE effects can partially reduce the observed difference \citep[e.g.][]{Potravnov_2023}. In the case of HD~152564, the NLTE abundance corrections for silicon and iron appeared to be small (see Sect. \ref{sect:lte_abund}), indicating a distinct vertical elemental stratification in the atmosphere. 

The depletion of the light elements (\ion{He}{}, \ion{O}{}, \ion{Mg}{}) observed in HD~152564 is in the qualitative agreement with results of the diffusion calculations. The mean iron overabundance shows satisfactory agreement between theory and observations. However, in a given temperature domain (around \te=12000~K) neither old equilibrium diffusion models by \citet{LeBlanc_2009} nor more sophisticated time-dependent ones by \citet{Alecian_2015} could reproduce the silicon overabundance over the entire surface and predict such strong vertical gradients of silicon and iron as shown by observations.

The use of spectral lines with different excitation energies allows us to probe the horizontal distribution of silicon at different depths in the atmosphere. The similarity of the abundance maps obtained using e.g. \ion{Si}{II} 5041~\AA\ and  \ion{Si}{II} 6660, 6665, 6671~\AA\, lines indicates that horizontal stratification occurs at certain depth inside the star and is then replicate to other layers through vertical diffusion. The latter is supported by the differences in the mean abundance values inferred from the lines with different formation depths.

\section{Conclusions}
\begin{enumerate}
 \item Based on the fitting of the observed \SED\ and the hydrogen line profiles we found atmospheric parameters of HD~152564 to be \te=11950$\pm$200 K, \logg=3.6$\pm$0.2 and the projected rotational velocity \vsini=69$\pm$2 \kms.\\
 \item The chemical composition of HD~152564 is typical for Ap stars and is characterised by a moderate deficiency of helium and some other light elements (\ion{O}{}, \ion{Mg}{}) and pronounced excesses of all measured heavier elements with atomic numbers up to 26. Some elements, namely \ion{Si}{} and \ion{Fe}{} show clear evidence of vertical stratification in the HD~152564 atmosphere. \\ 
 \item Both photometric and line profile variability of HD~152564 are consistent with the oblique rotator model with the inclination angle $i=51^{\circ}$ and inhomogeneous horizontal distribution of elements in the atmosphere.\\
 \item With the Doppler Imaging technique we investigated surface abundance distribution for 5 elements. The investigated elements showed variety of surface distributions: equatorial sequence of spots (\ion{Si}{},\ion{O}{}), circumpolar rings (\ion{Fe}{},\ion{Mg}{}), and high-latitude spots (\ion{He}{}) - well within the diversity observed in other magnetic CP stars.\\   
 \item Silicon abundance maps as traced by the lines from different ionisation stages and with different excitation energies turned out to be geometrically identical. This confirms the same horizontal distribution of silicon in the atmosphere at different depths. However, different mean abundances derived from moderate- and high-excitation lines indicate the presence of the vertical silicon abundance gradient in the atmosphere.\\.
 \item The maximum on the photometric light curve coincides with the passage through the central meridian of the most overabundant silicon spots.\\
 \item In spite of the significant (for Ap star) axial rotation, selective atomic diffusion has built up prominent vertical and horizontal abundance gradients in the  atmosphere of HD~152564. The three-dimensional distribution of abundances as well as the NLTE effects should be further taken into account for rigorous modelling of the parameters of the emergent radiation of HD~152564.
\end{enumerate}

\section*{Acknowledgements}

All observations reported in this paper were obtained with observational programs 2018-1-MLT-008 and 
2020-1-MLT-003 at the Southern African Large Telescope (SALT) 
supported by the National Research Foundation (NRF) of South Africa.
AK acknowledges the Ministry of Science and Higher Education of the Russian Federation grant 075-15-2022-262 (13.MNPMU.21.0003).\\
We acknowledge L. Mashonkina, A. Romanovskaya and T. Sitnova for their help in NLTE calculations.

\section*{DATA AVAILABILITY}

The data underlying this article will be shared on reasonable request to the corresponding author.




\bibliographystyle{mnras}
\bibliography{references.bib} 



\appendix
\section{List of spectral lines}
\begin{table*}
	\caption{List of spectral lines and the individual abundances in HD~152564.}
	\label{lines}
	\tabcolsep=2pt
	\begin{tabular}{llrrc p{3cm} l llrrcl}
		\hline
		Element & Line, \AA\, & $E_i$, eV & $\log gf$ & $\log (A)_X$ & Reference, gf & Element & Line, \AA\, & $E_i$, eV & $\log gf$ & $\log (A)_X$ & Reference, gf\\
		\hline
		\hline
\ion{He}{I}   & 4026.185: & 20.964 & -2.620 &    --- &    WSG &  \ion{Si}{II}  & 5056.316: & 10.073 & -0.359 &    --- &    S-G \\  
\ion{He}{I}   & 4026.187: & 20.964 & -1.450 &    --- &    WSG &  \ion{Si}{II}  & 5466.461  & 12.525 & -0.311 &  -3.29 &    K14 \\  
\ion{He}{I}   & 4026.187: & 20.964 & -0.700 &  -1.66 &    WSG &  \ion{Si}{II}  & 5466.849  & 12.525 & -1.457 &    --- &    K14 \\
\ion{He}{I}   & 4026.198: & 20.964 & -1.450 &    --- &    WSG &  \ion{Si}{II}  & 5466.894  & 12.525 & -0.156 &    --- &    K14 \\
\ion{He}{I}   & 4026.199: & 20.964 & -0.970 &    --- &    WSG &  \ion{Si}{II}  & 5469.232: & 16.726 &  0.242 &  -3.34 &    K14 \\
\ion{He}{I}   & 4026.358: & 20.964 & -1.320 &    --- &    WSG &  \ion{Si}{II}  & 5469.451: & 12.880 & -0.853 &    --- &    K14 \\
\ion{He}{I}   & 4143.761  & 21.218 & -1.195 &  -1.70 &    WSG &  \ion{Si}{II}  & 5469.469: & 12.880 & -1.807 &    --- &    K14 \\
\ion{He}{I}   & 4387.929  & 21.218 & -0.883 &  -1.51 &    WSG &  \ion{Si}{II}  & 5867.427: & 14.503 & -0.043 &    --- &    K14 \\
\ion{He}{I}   & 4471.469: & 20.964 & -2.198 &    --- &    WSG &  \ion{Si}{II}  & 5868.444: & 14.528 &  0.410 &  -2.53 &    Bar \\
\ion{He}{I}   & 4471.473: & 20.964 & -1.028 &    --- &    WSG &  \ion{Si}{II}  & 6239.613: & 12.839 &  0.185 &  -3.95 &    K14 \\
\ion{He}{I}   & 4471.473: & 20.964 & -0.278 &  -1.67 &    WSG &  \ion{Si}{II}  & 6239.613: & 12.839 & -1.359 &    --- &    K14 \\
\ion{He}{I}   & 4471.485: & 20.964 & -1.028 &    --- &    WSG &  \ion{Si}{II}  & 6239.664: & 12.839 &  0.072 &    --- &    K14 \\
\ion{He}{I}   & 4471.488: & 20.964 & -0.548 &    --- &    WSG &  \ion{Si}{II}  & 6347.108  &  8.121 &  0.170 &  -3.43 &  VALD3 \\
\ion{He}{I}   & 4471.682: & 20.964 & -0.898 &    --- &    WSG &  \ion{Si}{II}  & 6371.371  &  8.121 & -0.040 &  -3.58 &  VALD3 \\
\ion{He}{I}   & 5875.599: & 20.964 & -1.511 &    --- &    WSG &  \ion{Si}{II}  & 6660.532  & 14.503 &  0.162 &  -2.22 & NIST22 \\
\ion{He}{I}   & 5875.614: & 20.964 & -0.341 &    --- &    WSG &  \ion{Si}{II}  & 6665.026  & 14.489 & -0.240 &  -2.93 & NIST22 \\
\ion{He}{I}   & 5875.615: & 20.964 &  0.409 &  -1.48 &    WSG &  \ion{Si}{II}  & 6671.841  & 14.528 &  0.410 &  -2.96 &    BBC \\
\ion{He}{I}   & 5875.625: & 20.964 & -0.341 &    --- &    WSG &  \ion{Si}{II}  & 7848.816: & 12.525 &  0.316 &  -3.81 & NIST22 \\
\ion{He}{I}   & 5875.640: & 20.964 &  0.139 &    --- &    WSG &  \ion{Si}{II}  & 7849.617: & 12.525 & -0.831 &    --- & NIST22 \\
\ion{He}{I}   & 5875.966: & 20.964 & -0.211 &    --- &    WSG &  \ion{Si}{II}  & 7849.722: & 12.525 &  0.470 &    --- & NIST22 \\
\ion{He}{I}   & 6678.154  & 21.218 &  0.329 &  -1.63 &    WSG &  \ion{Si}{III} & 4552.622  & 19.016 &  0.292 &  -2.46 & NIST22 \\
\ion{C}{II}   & 4267.000: & 18.045 &  0.563 &  -2.65 & NIST10 &  \ion{Si}{III} & 4567.840  & 19.016 &  0.070 &  -2.70 & NIST22 \\
\ion{C}{II}   & 4267.261: & 18.046 &  0.716 &    --- & NIST10 &  \ion{Si}{III} & 4574.757  & 19.016 & -0.410 &  -2.48 & NIST22 \\
\ion{C}{II}   & 4267.261: & 18.046 & -0.584 &    --- & NIST10 &  \ion{Ti}{II}  & 4464.449  &  1.161 & -1.810 &  -6.11 &    PTP \\
\ion{C}{II}   & 6578.050  & 14.448 & -0.021 &  -2.72 & NIST10 &  \ion{Ti}{II}  & 4468.492  &  1.130 & -0.630 &  -6.11 &   WLSC \\
\ion{C}{II}   & 6582.880  & 14.448 & -0.323 &  -2.73 & NIST10 &  \ion{Ti}{II}  & 4501.258  &  1.115 & -0.770 &  -6.12 &   WLSC \\
\ion{O}{I}    & 6155.961: & 10.740 & -1.363 &    --- & NIST10 &  \ion{Ti}{II}  & 4563.752  &  1.221 & -0.690 &  -6.36 &    PTP \\
\ion{O}{I}    & 6155.971: & 10.740 & -1.011 &    --- & NIST10 &  \ion{Ti}{II}  & 4571.959  &  1.571 & -0.310 &  -5.96 &   WLSC \\
\ion{O}{I}    & 6155.989: & 10.740 & -1.120 &    --- & NIST10 &  \ion{Cr}{II}  & 4145.778  &  5.318 & -0.902 &  -5.35 &   GNEL \\
\ion{O}{I}    & 6156.737: & 10.740 & -1.487 &    --- & NIST10 &  \ion{Cr}{II}  & 4558.650: &  4.073 & -0.449 &  -5.57 &   PGBH \\
\ion{O}{I}    & 6156.755: & 10.740 & -0.898 &    --- & NIST10 &  \ion{Cr}{II}  & 4558.783: &  4.073 & -2.530 &    --- &    SLd \\
\ion{O}{I}    & 6156.778: & 10.740 & -0.694 &    --- & NIST10 &  \ion{Cr}{II}  & 4565.740  &  4.042 & -1.820 &  -5.60 &    SLd \\
\ion{O}{I}    & 6158.149: & 10.740 & -1.841 &    --- & NIST10 &  \ion{Cr}{II}  & 4588.199  &  4.071 & -0.627 &  -5.49 &   PGBH \\
\ion{O}{I}    & 6158.172: & 10.740 & -0.995 &    --- & NIST10 &  \ion{Cr}{II}  & 4592.049  &  4.073 & -1.221 &  -5.73 &   PGBH \\
\ion{O}{I}    & 6158.187: & 10.740 & -0.409 &  -3.93 & NIST10 &  \ion{Cr}{II}  & 4616.629  &  4.072 & -1.361 &  -5.55 &   PGBH \\
\ion{O}{I}    & 7771.944: &  9.146 &  0.369 &    --- & NIST10 &  \ion{Cr}{II}  & 4618.803  &  4.073 & -0.840 &  -5.25 &    SLd \\
\ion{O}{I}    & 7774.166: &  9.146 &  0.223 &    --- & NIST10 &  \ion{Fe}{I}   & 4045.812  &  1.485 & -2.700 &  -4.53 &    FMW \\
\ion{O}{I}    & 7775.388: &  9.146 &  0.002 &  -4.13 & NIST10 &  \ion{Fe}{I}   & 4063.594  &  1.557 & -2.420 &  -4.28 &    FMW \\
\ion{Mg}{II}  & 4481.126: &  8.863 &  0.740 &  -5.04 &     KP &  \ion{Fe}{I}   & 4071.737  &  1.608 & -2.169 &  -4.85 &    FMW \\
\ion{Mg}{II}  & 4481.150: &  8.863 & -0.560 &    --- &     KP &  \ion{Fe}{I}   & 4383.544  &  1.485 & -2.976 &  -4.34 &    FMW \\
\ion{Mg}{II}  & 4481.325: &  8.863 &  0.590 &    --- &     KP &  \ion{Fe}{II}  & 4491.397  &  2.855 & -2.700 &  -3.90 &     KK \\
\ion{Mg}{II}  & 4384.637  &  9.995 & -0.790 &  -4.99 &     KP &  \ion{Fe}{II}  & 4508.280  &  2.855 & -2.420 &  -3.80 & D-HLSC \\
\ion{Mg}{II}  & 4390.514: &  9.999 & -1.490 &    --- &     KP &  \ion{Fe}{II}  & 4522.627  &  2.844 & -2.169 &  -3.92 &     RU \\
\ion{Mg}{II}  & 4390.572: &  9.999 & -0.530 &  -4.99 &     KP &  \ion{Fe}{II}  & 4576.332  &  2.844 & -2.976 &  -4.10 &     RU \\
\ion{Mg}{II}  & 4427.994  &  9.995 & -1.210 &  -4.86 &     KP &  \ion{Fe}{II}  & 4598.485  &  7.804 & -1.536 &  -3.86 &     RU \\
\ion{Al}{II}  & 4663.046  & 10.598 & -0.284 &  -6.58 &    WSM &  \ion{Fe}{II}  & 4640.803  &  7.708 & -1.737 &  -3.97 &     RU \\
\ion{Al}{II}  & 6243.367  & 13.076 &  0.680 &  -6.85 &    WSM &  \ion{Fe}{II}  & 4731.447  &  2.891 & -3.127 &  -4.08 &     RU \\
\ion{Si}{II}  & 4128.053  &  9.836 &  0.410 &  -3.84 &    S-G &  \ion{Fe}{II}  & 5018.435  &  2.891 & -1.399 &  -3.87 &    K13 \\
\ion{Si}{II}  & 4130.871: &  9.838 & -0.824 &    --- &   BBCB &  \ion{Fe}{II}  & 5026.797  & 10.307 & -0.214 &  -3.93 &    K13 \\
\ion{Si}{II}  & 4130.893: &  9.838 &  0.530 &  -3.75 &    S-G &  \ion{Fe}{II}  & 5045.106  & 10.307 & -0.151 &  -3.77 &    K13 \\
\ion{Si}{II}  & 4621.418: & 12.525 & -0.540 &  -3.49 & NIST22 &  \ion{Fe}{II}  & 5067.890  & 10.328 & -0.078 &  -3.65 &     RU \\
\ion{Si}{II}  & 4621.695: & 12.525 & -1.680 &    --- & NIST22 &  \ion{Fe}{II}  & 5291.660  & 10.480 &  0.561 &  -3.71 &    K13 \\
\ion{Si}{II}  & 4621.721: & 12.525 & -0.380 &    --- & NIST22 &  \ion{Fe}{II}  & 5325.552  &  3.221 & -3.185 &  -3.72 &    K13 \\
\ion{Si}{II}  & 5041.023  & 10.066 &  0.150 &  -3.48 &  VALD3 &  \ion{Fe}{III} & 4382.501  &  8.247 & -2.980 &  -2.78 &    K10 \\
\ion{Si}{II}  & 5055.984: & 10.073 &  0.530 &  -3.81 &  VALD3 &  \ion{Fe}{III} & 4395.751  &  8.256 & -2.586 &  -3.18 &    K10 \\
		\hline
	\end{tabular}

		{\it Note.}	Close blends are marked by colon at wavelength, abundance is given at the strongest blend component. WSG = \citet{WSG}; NIST10 = \citet{NIST10};
	KP = \citet{KP}; WSM = \citet{WSM}; S-G = \citet{S-G}; BBCB = \citet{BBCB}; Bar = \citet{Bar}; NIST22 = \citet{NIST22}; BBC = \citet{BBC}; PTP = \citet{PTP}; 
	WLSC = \citet{WLSC}; GNEL = \citet{GNEL}; PGBH = \citet{PGBH}; SLd = \citet{SLd}; KK = \citet{KK}; FMW = \citet{FMW}; D-HLSC = \citet{D-HLSC}; RU = \citet{RU}; K10 = \citet{K10};
	K13 = \citet{K13}; K14 = \citet{K14}; VALD3 = \citet{Ryabchikova_2015}.
\end{table*}

\bsp	
\label{lastpage}

\bsp	
\label{lastpage}
\end{document}